\documentstyle[preprint,aps]{revtex}
\begin{document}
\draft
%\twocolumn[\hsize\textwidth\columnwidth\hsize\csname @twocolumnfalse\endcsname

\title{Spatio-temporal dynamics and plastic flow of vortices 
in superconductors with periodic arrays of pinning sites} 
\author{C.~Reichhardt, J.~Groth, C.~J.~Olson, Stuart Field, and 
Franco Nori}
\address{Department of Physics, The University of Michigan, 
Ann Arbor, Michigan 48109-1120}

\date{\today}
\maketitle
\begin{abstract}
We present simulations of flux-gradient-driven superconducting rigid vortices 
interacting with square and triangular arrays of columnar pinning sites in an 
increasing external magnetic field. These simulations allow us to 
quantitatively relate spatio-temporal microscopic information of the 
vortex lattice with typically measured macroscopic quantities, such 
as the magnetization $M(H)$. 
The flux  lattice does not become completely commensurate with the pinning
sites 
throughout the sample at the magnetization matching peaks, but forms a 
commensurate lattice in a region close to the edge of the sample. 
Matching fields related to unstable vortex configurations do not produce 
peaks in $M(H)$.  
We observe a variety of evolving complex flux profiles, including flat 
 terraces or
plateaus separated by winding current-carrying strings and,
near the peaks in $M(H)$, plateaus only in certain 
regions, which move through the sample as the field increases.
\end{abstract} 
\vskip1pc  
%%\vskip2pc]  
\narrowtext

\section{Introduction} 

Flux pinning in type II superconductors is of both technological and 
scientific interest.  While most experiments focus on the effects of 
random pinning distributions, some investigations have been carried out 
on periodic arrays of pinning sites (PAPS) \cite{reviews}.
These find striking peaks in the magnetization
\cite{L,Metulshko,Jonkeree,Baert} 
and critical current $ J_{c} $. 
These peaks are believed to arise from the greatly enhanced pinning 
that occurs when parts of the vortex lattice (VL) become commensurate 
with (i.e., match) the underlying PAPS.   Under such conditions, 
high-stability vortex configurations are produced 
which persist under an increasing current or external field.
Other important vortex matching effects have also recently been 
observed in a variety of different superconducting systems 
\cite{Ovch,hunnekes,JJ,networks,imaging,oussena,Grishin},
including Josephson junctions\cite{JJ}, 
superconducting networks\cite{networks,imaging}, 
and the remarkable matching of the VL to the crystal structure of 
YBa$_2$Cu$_3$O$_7$ due to intrinsic pinning\cite{oussena}.  

Commensurability effects are also important in superconducting
systems with {\it random\/} arrays of columnar pinning sites,
where a Mott insulator phase is predicted \cite{Nelson}
at the matching field, 
where the number of vortices equals the number of pinning sites.
In this phase the vortex lattice locks into the
pinning array preventing further vortices from entering the sample 
and creating a Meissner-like effect. 

Non-superconducting systems also exhibit magnetic-field-tuned matching 
effects, notably in relation to electron motion in periodic structures
where unusual behaviors arise due to the incommensurability of the
magnetic length with the lattice spacing\cite{azbel}.
A recent example of such a system is provided by the anomalous Hall 
plateaus of ``electron  pinball''\cite{pinball} orbits scattering 
from a regular array of antidots.
Commensurate effects also play central roles in many other 
areas of physics, including plasmas, nonlinear dynamics\cite{NL}, 
the growth of crystal surfaces, domain walls in incommensurate solids, 
quasicrystals, Wigner crystals, as well as spin and charge density 
waves\cite{CDW}.

To investigate the microscopic vortex dynamics in systems with a PAPS we
have performed extensive molecular dynamics simulations using a wide
variety of relevant parameters which are difficult to {\it continuously\/}
tune experimentally, such as disorder in the pinning lattice
geometry, sample size, vortex density $n_v$, as well as pinning density
$ n_{p}$, radius $\xi_{p}$, and maximum pinning force (or strength) $f_{p}$.
We find a rich variety of behavior in which all these parameters play an
important role.

Recent experiments and theories (see, e.g., Refs [1-8])
involving  periodic and commensurate pinning effects in superconductors 
have raised questions that can be systematically explored 
by our computer simulations. 
One such question addressed in this work involves the topological ordering 
and dynamics of the vortices at commensurate and incommensurate 
fields and how this {\it microscopic\/} order relates to bulk 
{\it macroscopic\/} experimentally measurable quantities,
such as the magnetization $M(H)$.
At the matching fields, the vortex lattice is expected to be 
topologically ordered; however, most experiments 
have not been able to directly image the vortex lattice to determine 
the precise vortex arrangements for varying external magnetic fields. 
In particular, at the matching field most experiments cannot directly 
discern whether several vortices sit at each pinning site or 
the vortices sit at the interstitials (regions between pinning sites).
{\it Interstitially\/} pinned vortices are those which are trapped 
by other vortices, and not by defects.
Interstitial pinning refers to ``magnetic caging", in contrast to 
the standard, and stronger, ``core pinning" by defects.
Our direct knowledge of the vortex positions allows us to 
determine the degree of topological disorder in the vortex 
lattice at the incommensurate fields and compare it with 
the ordered phases at the matching fields. 

Another important question that can be directly studied 
using computer simulations is the vortex lattice dynamics near 
and above the first matching field.  Recent theories 
\cite{IK,Leo} and experiments 
\cite{Majer,Welp,Rosenbaum,Rosseel} suggest that beyond the first 
matching field, weakly pinned interstitial vortices should be
more mobile than vortices at the defect sites; 
thus, quantities such as critical currents and magnetization 
should drop considerably. Since these interstitial vortices 
should be flowing around the strongly pinned immobile vortices,
{\it plastic} flow should occur in these systems.    
Plastic flow of vortices in superconductors has recently attracted 
considerable attention \cite{Science,Tonomura,Yaron} 
and superconductors with a PAPS are an ideal system in which to study it.  
Our simulations allow us to explicitly show, in a quantitative manner,
how dominant interstitial vortex motion is beyond the matching $B_{\phi}$. 
Direct imaging of the vortex lattice can also determine exactly how 
different periodicities in the underlying pinning geometry affect 
the ordering of the vortex lattice.    

It is important to emphasize that our simulations take into account 
the critical state, \cite{Bean,Kim}, in which a {\it gradient\/} in 
the flux profile arises as flux penetrates a superconductor. 
Recently, Cooley and Grishin \cite{Grishin} presented a 1D static model 
of the Bean critical state with a PAPS, in which a terraced flux 
profile arises.  
In this work we will focus on a 2D {\it dynamical\/} model of 
flux-gradient-driven  vortices;  this 2D dynamical descripton produces 
a far more complex and richer behavior than that predicted by 
1D static models \cite{Grishin}. 

Section II describes the numerical algorithm used here and compares 
our system to other superconducting structures with periodic pinning. 
In section III we present calculated values for the magnetization 
$ M(H(t)) $ for superconducting samples with either a square or a 
triangular array of pinning sites.   This computed macroscopic average, 
$M(H)$, is then related to microscopic quantities which describe the 
evolving topological order of the vortex lattice (e.g., the density of 
$n$-coordinated defects $P_n(H)$ in the flux lattice).
Section IV shows specific vortex configurations and describes 
their dynamics close to the first commensurate peak in $M(H)$.
Section V presents vortex lattices at high matching field values 
and compares them with the low matching field configurations. 
Section VI shows the effects on the magnetization of varying the 
pinning strength, density, and disorder in the pinning lattice. 
Section VII contains a summary of our results and a discussion of 
how they compare to recent theories and experiments.
   
\section{Simulation} 

We consider a 2D transverse slice (in the $x$--$y$ plane) of an infinite 3D 
slab at zero temperature containing rigid vortices and columnar defects, 
all parallel to both the sample edge and the applied field 
${\bf H}=H{\bf {\hat z}}$.  
Pinning is placed in the central $ 75\% $ of the system (which is our
actual sample) with the outer $ 25\% $ free of pinning sites. We simulate 
the ramping of an external field as described in 
\cite{Richardson,Reichhardt} by the slow addition of vortices to the 
unpinned region. These vortices attain a uniform density $n$ allowing us 
to define the external field $H = n\Phi_0$, where $ \Phi_0 = hc/2e$.
 As the external density of the flux lines builds, they will, 
through their own repulsion, force their way into the sample 
where they interact with the lattice of columnar defects. 
We correctly model the vortex-vortex force by a modified Bessel 
function $ K_{1}(r/\lambda)$, where $\lambda$ is the penetration depth.
For the vortex-pin interactions, we assume the 
pinning potential well to be parabolic. For the simulations presented 
here, the pinning range (i.e., the radius of the parabolic well) 
is $\xi_p= 0.25\lambda $, 
far less than our vortex-vortex interaction cutoff of $ 6\lambda $. 
The overdamped equation of motion is
$$ {\bf f}_{i} = {\bf f}_{i}^{vv} + {\bf f}_{i}^{vp} = \eta{\bf v}_{i} \ , $$
where the total force $ {\bf f}_{i} $ on vortex $ i $ (due to the other 
vortices $ {\bf f}_{i}^{vv}$, and pinning sites $ {\bf f}_{i}^{vp} $) 
is given by 
%  This two-line Eq. is OK for SHORT PREPRINT format
\begin{eqnarray} 
{\bf f}_{i} & = &  \ \sum_{j=1}^{N_{v}}\, f_{0} \,  K_{1} 
\left( \frac{ |{\bf r}_{i} - {\bf r}_{j}| }{ \lambda} \right) 
\, {\bf {\hat r}}_{ij}  \nonumber 
         \\ & + & \sum_{k =1}^{N_{p}} \frac{f_{p}}{\xi_{p}} \ 
|{\bf r}_{i} - {\bf r}_{k}^{p}| \ \Theta\left( 
\frac{ \xi_{p} - | {\bf r}_{i} - {\bf r}_{k}^{p} | }{\lambda}  \right) \ 
{\bf {\hat r}}_{ik} \, .
\end{eqnarray}
%  This one-line Eq. is OK for LONG PREPRINT format
%\begin{equation} 
%$
%{\bf f}_{i} = \, \sum_{j=1}^{N_{v}}\, f_{0} \,  K_{1} \!
%\left( \frac{ |{\bf r}_{i} - {\bf r}_{j}| }{ \lambda} \right) 
%      (        |{\bf r}_{i} - {\bf r}_{j}| /  \lambda        ) 
%\, {\bf {\hat r}}_{ij}  
%         + \sum_{k =1}^{N_{p}}  ( f_{p} / \xi_{p} ) % \times
%|{\bf r}_{i} - {\bf r}_{k}^{p}| \ \Theta  \!  
%\left( \frac{ \xi_{p} - |{\bf r}_{i}-{\bf r}_{k}^{p} | }{\lambda} \right) \ 
%      [      ( \xi_{p} - |{\bf r}_{i}-{\bf r}_{k}^{p} | )/\lambda  ]
%{\bf {\hat r}}_{ik} \, .
%%\end{equation}
%$
%\noindent 
Here $ \Theta $ is the Heaviside step function, 
$ {\bf r}_{i}$  $ ({\bf v}_{i}) $
is the location (velocity) of the $i$th vortex, 
$ {\bf r}_{k}^{p} $ 
is the location of the $k$th pinning site, 
$ {\bf {\hat r}}_{ij} = ({\bf r}_{i}-{\bf r}_{j}      )/
|{\bf r}_{i}-{\bf r}_{j}| $, 
$ {\bf {\hat r}}_{ik} = ({\bf r}_{i}-{\bf r}_{k}^{p})/
|{\bf r}_{i}-{\bf r}^{p}_{k}| $,
$f_p$ is the maximum pinning force (or strength),
and we take $\eta=1$.
Unlike previous simulations which have investigated systems with 
random pinning arrays \cite{Richardson,Reichhardt,Olson,simulations},
here we place the pinning sites in square as well as triangular arrays.

We measure all forces in units of $ f_{0} = \Phi_{0}^{2}/8\pi^{2}\lambda^{3}$,
fields in units of $\Phi_0/\lambda^2$, and lengths in units of 
$\lambda$.
Our simulation correctly models the driving force as a result of 
{\it local\/} interactions; no artificial ``uniform" external 
force is applied to the flux lines.  
As the vortices enter the sample (i.e., the pinned region) we can compute
experimentally measurable quantities, including the density of flux lines 
$B(x,y,H(t))$, and the local $J_c(x,y,H(t))$ and global currents 
$J_c(H(t))$. 
Here, we will mainly focus on the most commonly measured macroscopic quantity,
the global magnetization 
$$M = (4\pi V)^{-1}\int (H - B) \; dV \ , $$
and relate it to the spatio-temporal dynamics of the vortices.
 
Our simulation procedure and the samples used here have several 
differences that distinguish them from recent experiments.
First, our samples have a slab geometry, unlike the films used in 
\cite{Metulshko,Jonkeree,Baert}, where 
demagnetization effects are important.
Second, the size and strength of our pinning sites are 
much smaller than those considered in 
\cite{Metulshko,Jonkeree,Baert,Ovch}.  Thus
at higher matching fields, interstitial pinning will be important 
rather than multiple vortices per pinning site, 
as in the experiments \cite{Metulshko,Jonkeree,Baert,Ovch}.
Moreover, the relatively small radius of our pinning sites, among
other reasons, makes our system very different from wire networks, 
where interstitial vortices cannot exist.    
Finally, and most importantly, our simulations are {\it dynamical}. 
The vortices are driven by the flux gradient as the external magnetic 
field is increased, and we can observe how the vortices are moving at 
the various external field values. This is considerably different from 
the static experimental results where the vortex motion has not been
imaged. 

\section{ Relation between macroscopic averages and the microscopic
topological order of individual flux lines}

Previous simulations of {\it flux-gradient-driven\/} 
vortices conducted in 1D \cite{Richardson} and 2D \cite{Reichhardt}  
with dense randomly placed pins produced a modified Bean-Kim model in 
which the flux profile had no terraces and the magnetization curves 
were very similar to those observed by standard experiments. 
However, with a PAPS, a very different behavior arises. 
Figure 1(a) presents the first quarter of the initial magnetization curve for 
a square PAPS. 
Although our simulations produce complete hysteresis loops 
\cite{Reichhardt,Olson}, the ramp-up phase contains the relevant information.  
As the external field $H$ is increased from zero, 
the flux fronts move towards the center of the sample and $M(H)$ rises; 
when 
the two flux fronts meet at the center of the sample, $M(H)$ peaks.  
This initial peak is thus unrelated to 
the PAPS, and is seen even in random arrays\cite{Reichhardt}.
However, as the field is increased further, we find 
three cusps near the matching fields (MFs) given by 
$ n_{v}/n_{p} \approx 1/2$, $1$, and $2$, with the peak at 
$ n_{v}/n_{p} \approx 3$ missing  
since the VL cannot form a stable configuration commensurate with the
underlying square geometry for $ n_{v}/n_{p} \approx 3 $.
The VL can form a stable configuration commensurate with the
underlying square geometry for $ n_{v}/n_{p} \approx 2 $.
 The peaks occur slightly {\it after\/} the exact matching due to the 
flux gradient in the sample.  
For the range of fields studied here ($ n_{v}/n_{p} \lesssim 4 $),
we observe that the radius of the pinning wells $\xi_p$ is too small 
to permit multiple vortices per well; 
thus, the higher MFs form lattices with vortices pinned at 
interstitial sites.  
In flux creep experiments with a square PAPS\cite{Metulshko} 
it was suggested that multiple vortices were trapped in a single 
large well at the $n_{v}/n_{p} \approx 3 $ MF.  
Indeed, our results indicate that for a square PAPS any experimentally 
observed \cite{Metulshko} MFs near $n_{v}/n_{p} \approx 3 $ 
must be due to {\it multiple\/} vortices per well.

In order to correlate the behavior of 
this {\it macroscopic quantity\/} $M(H)$ 
with  the actual {\it microscopic dynamics\/} of individual vortices, it is 
useful to monitor the field-dependent fraction (or probability) $P_n(H)$ of 
vortices with coordination number $n$.
This quantity is obtained from the Voronoi (or generalized Wigner-Seitz)
construction for the individual vortices and is shown in Fig.~1(b--e, g--j).
The lengths of each side of a given Voronoi cell are compared, and
if any side is less than one-tenth of the lengths of the other sides, it 
is ignored. This permits us to examine square lattices, which can
be difficult to deal with using standard Voronoi algorithms. 
In the absence of any pinning, the flux lines form a perfectly hexagonal
lattice ($P_6(H) = 1$); the presence of pinning sites introduces disorder
into the flux lattice. 
The fraction of $4$--fold-coordinated vortices $P_4(H)$, in Fig.~1(b), 
quantifies the degree of matching of the vortex lattice with the square 
pinning array 
at the commensurate fields $n_{v}/n_{p} \approx 1/2$, $1$, and $2$.  
At these fields where $P_4$ peaks, {\it all\/} the other 
probabilities show a clear dip.
Notice also the increase in $P_4(H)$ near the $1/4$ and $1/8$ rational 
matching fields; these are not obviously reflected in $M(H)$, due to 
the usual large peak at $H\approx H^*$.  
Moreover, rational matching peaks higher than one (e.g., $3/2$, $5/2$) 
are not observed, due to the presence of a flux gradient.
$P_6(H)$ greatly increases between the MFs, since there the 
vortex-vortex forces dominate over the vortex-pinning interactions.
For very weak fields, the low numbers of vortices makes the Voronoi 
construction ill-defined.

Figure 1(f) shows $M(H)$ for a triangular PAPS.
Here the MFs at $ n_{v}/n_{p} \approx 1 $ and $3$ 
produce peaks in $ M(H) $. The $ n_{v}/n_{p} \approx 2 $ peak is missing 
for this weak pinning case 
(although we observe a small peak for strong pinning $f_p=2.0f_0$) 
because this vortex configuration is less stable than the triangular 
VL at the same $n_v$.
Figures 1(g--j) show the corresponding fraction of $n$-fold coordinated 
vortices.  
Here $P_6(H)$ increases at the MFs $1$ and $3$,
and also has a smaller increase near $1/2$.
Notice that $P_4(H)$, $P_5(H)$, and $P_7(H)$ decrease at the 
MFs $1$ and $3$.
$P_6(H)$ tends to increase with field for both square and triangular 
PAPS.  Since here the $5$-$7$ disclinations are paired, $P_5(H)$ and 
$P_7(H)$ follow each other closely.

Figure 2(a--b) [2(c--d)] presents the actual positions of the vortices
and the square [triangular] PAPS for a small part of the sample
used in Fig.~1(a) [1(f)] at the $n_v/n_p \approx 1, 2 \ [1, 3]$ MFs. 
Due to the field gradient noticeable here, 
the VL is only commensurate over this small range.
In Fig.~2(a) all the vortices are pinned at the defect sites; 
these vortices are said to experience ``core pinning", instead
of ``magnetic pinning".
The additional vortices that enter the sample when the external field
is increased will not be trapped by the already-occupied pinning sites, 
but will still feel a magnetic ``caging" effect due to the repulsion 
produced by other vortices.  These are known as interstitially-pinned,
or ``magnetically-pinned", vortices.  
Here, we do not consider ``magnetic pinning" due to magnetic 
impurities in the sample (e.g., iron dots); 
vortices are pinned either by defects or by other pinned vortices.
In Fig.~2(b), half the vortices are interstitially pinned between the 
vortices located at the pinning sites, forming a square vortex lattice. 
In Fig.~2(c-d) the same cycle is observed only now there are two 
interstitially pinned vortices for every vortex trapped at a defect site. 

A recent experiment \cite{Baert} finds that the particular type 
of pinning geometry (i.e., square, triangular) produces little 
difference in the matching fields observed (i.e., peaks in $M(H)$
appear at {\it all integer} matching fields). 
This is in strong contrast to our results where the particular kind 
of pinning geometry clearly affects  where matching peaks occur.
Our results can be understood in terms of interstitial 
vortices being constrained to form a lattice with the same 
symmetry as the underlying pinning lattice. 
For a triangular array of weak pinning sites, a triangular vortex 
lattice with two vortices per pin, in which one vortex is in a 
pinning site and the other is interstitially confined, is unstable
as can be shown by a simple linear stability analysis.
Thus at the second matching fields, we do not observe a peak in 
either the $6$--fold coordination number or the magnetization.
Similarly, at the third matching field of a square array of pinning sites,
a square vortex lattice is unstable and a peak in the both the 
$4$--fold coordination number and the magnetization is not observed there. 
For the large and strong pinning sites used in certain experiments,
e.g., \cite{Baert}, however, the onset of multiple vortices per pinning site 
leads to reduced sensitivity to the underlying geometry of the 
pinning lattice.

\section{Vortex dynamics and plastic flow close to a commensurate field.}

To investigate further the VL spatio-temporal dynamics as $M(H)$ 
crosses the first commensurate ($ n_{v}/n_{p} \approx 1 $) peak 
for the system in Fig.~1(a), we present snapshots in Fig.~3(a--d) 
of the real-space time evolution of the VL Voronoi cells.
The locations of these snapshots on the magnetization curve 
are indicated in Fig.~1(a) by the letters a--d.
Figure 3(a) shows the flux distribution just before the 
$n_{v}/n_{p} \approx 1$ peak in $M(H)$.
Commensurability, as indicated by the presence of mostly square Voronoi cells,
starts at the edges of the sample as the magnetization begins to rise.
Figure 3(b) shows the flux distribution precisely at the magnetization peak.
The VL is strongly commensurate with the PAPS, but only in regions close 
to the sample {\it edges}.  
Beyond the peak, Fig.~3(c), the commensurability between the VL and 
the PAPS moves from the {\it edges\/} of the sample to the {\it center}, 
and the  edges start to become disordered as vortices move 
(by flowing between the vortices trapped at the pins) into the 
interstitial sites, indicated by the smaller (lighter) 
polygons oriented $45^{\circ}$ from the sample edge.
This clearly shows that plastic flow of interstitial vortices 
dominates vortex transport when $ n_{v} > n_{p} $. 
Finally, Fig.~3(d) shows the vortex configuration near the minimum 
between the two commensurate peaks, where the flux distribution 
is disordered throughout the sample.   
As the external field is further increased, the vortex lattice 
remains disordered until the second MF, 
where the cycle is repeated; however, this 
time the VL also includes interstitially-pinned vortices.

Since the commensurate portions of the vortex lattice are uniform in 
density, a terrace in the flux gradient arises
with large currents at its boundary, where the flux density abruptly
changes.  In larger systems, we observe a flux profile with several 
separate commensurate regions (plateaus or terraces) inside the sample.  
For increasing $H$, the commensurate portions or terraces 
gradually move towards the center of the sample, reaching the 
center, where they eventually disappear.
We point out that we do not assume {\it a priori\/} any particular form 
of the profile; we compute it dynamically by ramping up the external 
field $H$ from zero.  This dynamically-generated 2D terraced profile 
is somewhat similar to the static 1D one obtained in \cite{Grishin}.
However, while all integer MF's produced peaks in that 1D model,
in our 2D model, for a given pinning geometry, only certain MFs actually 
produce peaks in $M(H)$. 
We find, for instance, that a kagom\'e and honeycomb PAPS do 
{\it not\/} have a peak at $n_v/n_p \approx 1$ for weak pinning 
($f_p \approx 0.1 f_0 $); a 1D analysis cannot distinguish 
these important geometrical effects. 
We have also found that 
the terraced ordered regions of the VL 
(corresponding to consecutive matching fields that produce 
commensurability peaks in $M(H)$)
are separated by distinct domains where the flux lattice is disordered. 

\section {Complex patterns of magnetic flux: islands and striped domains}

Figures 4(a,b) present the Voronoi cells for a system with a lower pin 
density $ n_{p} = 0.49/\lambda^{2} $, but about four times higher pin 
strength $ f_{p} = 1.4f_{0} $, than the $n_{p}$ and $f_{p}$ used 
for Fig.~1(a). 
The magnetization (not shown) is initially very high 
when $ n_{v} < n_{p} $, because the strong pinning forces create a very 
high $ J_{c}$; however, after the first MF (i.e., when $n_v > n_p $), 
there is a very rapid drop off in $ M(H) $ and $ J_{c}(H) $.
This rapid drop occurs because the characteristic pinning mechanism 
changes from individual columnar defect pinning to much weaker 
interstitial pinning.  This result, obtained via a dynamical simulation, 
is consistent with previous work using a time-independent formalism 
\cite{IK,Leo}. At high fields intricate vortex patterns appear with 
{\it islands} and {\it stripes} even though no noticeable change in 
the magnetization can be seen. The extremely weak interstitial pinning 
found at these high fields means that ordered regions of vortices 
with different orientations are close enough energetically that 
different local configurations can coexist in the sample.  
Notice that, in general, the field distribution is {\it not\/} 1D terraced
as in \cite{Grishin}.  For instance, contours of constant $B =$  indicate 
that the ``islands'' and ``stripes'' in Figs.~4(a,b) have near-zero current 
($\nabla \times {\bf B} \approx 0$) and are surrounded by current-carrying 
{\it winding\/} strings composed mostly of small current loops. 
   
\section {Effects of the microscopic pinning parameters and 
controlled disorder on the magnetization. }

Submatching peaks 
[indicated by arrows at $n_{v}/n_{p} \approx 1/4$ and $1/2$ in Fig.~5(a)]
in $M(H)$ become evident for samples with a large density of weak 
pinning sites. 
At the half matching field a portion of the vortices in the sample 
are arranged in a checkerboard pattern similar to that seen in 
wire networks \cite{imaging};
however, submatching configurations
at $ n_{v}/n_{p} = 1/3$, $2/5$, $2/3$, and $3/4$, 
which are found for wire networks,  
are not observed here. Our result for the submatching configurations   
are in agreement with the experiments in Ref.~\cite{Metulshko},
which use a system resembling ours. 
As the pinning force is increased the width of $M(H)$ increases,
making it difficult to identify the submatching fields below $1/2$.
As the flux gradient increases, the submatching vortex configurations 
can only form on smaller regions of the sample.
It is thus easier to observe submatching field configurations 
in a sample with a higher density ($n_p=1.83/\lambda^2$)
of weak pinning sites [e.g., Fig.~5(a)] 
than in a sample with a lower density ($n_p=0.86/\lambda^2$) 
of pins [see Fig.~5(b)].
This is the reason why experiments (e.g., Ref.~\cite{Metulshko}) 
can distinguish these peaks in $M(H)$ more clearly when the 
temperature is near $T_c$, where pinning is weak and submatching vortex 
configurations can occupy a larger portion of the sample.

Figure 5(b) shows $M(H)$ for maximum pinning forces between 
$f_{p} =0.1f_{0}\;$ and $f_{p} =0.4f_{0}$, with all other parameters 
fixed and ($n_p=0.86/\lambda^2$).  
The overall magnetization width increases and the peaks 
at $n_{v}/n_{p} \approx 1$ and $2$ [indicated by the arrows in 
Figs.~5(b,c)] have sharper drop offs right after the MFs.
The sharp drop off after the first matching peak 
can be understood as a crossover in the pinning 
mechanism from  vortices strongly pinned at defect sites 
to much weaker interstitial pinning. Such a crossover has been 
considered theoretically \cite{IK,Leo,Olson} and 
observed experimentally \cite{Majer,Welp,Rosenbaum,Rosseel} 
for random pinning arrays.  
Since $ J_{c} $ is directly related to the width of $M$, these
results indicate that enhancements of $ J_{c} $ may be restricted 
to fields less than the first matching field in the case
when only one vortex can be  trapped by each pinning site. 

To examine the effects of disorder we displaced the pins from their 
ordered positions in random directions up to a maximum
distance $ \delta r $. 
Figure 5(c) shows the effects on the first two peaks of $M$ of gradually
increasing the disorder in a square PAPS.  As the disorder is increased from 
$\delta r = 0 $ to $ \delta r = \lambda/8 $ (bottom two curves), 
the second peak at $n_{v}/n_{p} \approx  2 $ is suppressed, 
while the first peak remains. 
This occurs because the second peak is caused by {\it weaker\/} interstitial 
pinning---thus more susceptible to disorder. 
As the disorder further increases to $\delta r = \lambda/4$ (top curve), 
the first peak disappears as well.  
Wire networks \cite{networks} in which disorder 
is gradually introduced show similar effects since, in both cases,
commensurate peaks decrease with increasing disorder. 

\section{Summary and Discussion} 

In summary, we have presented simulations of flux-gradient-driven 
superconducting vortices interacting with various PAPS, on systems 
of varying pinning strength, density, geometry, and disorder.
The comparison between $M(H)$ and the $P_n(H)$s, shown in Fig.~1, 
quantitatively correlates how the geometry of the pinning lattice 
affects the underlying microscopic dynamics of individual vortices 
and its relation to macroscopic measurable quantities 
like $M(H)$ and $J_{c}$.

Our simulations explicitly show that at certain matching fields
a subset of the vortex lattice  becomes commensurate with the 
underlying array of pinning sites.
Because of the gradient in the flux density due to the critical state, 
the commensurate portions of the vortex lattice first start at the 
edge of the sample and then gradually move to the center.   
These configurations persist for a finite range of increasing 
external field, creating a peak in $M(H)$.  
By monitoring the vortex coordination number as a function of external field, 
$P_n(H)$, we have shown that the vortex lattice goes through a series 
of ordered-disordered states  
with the higher matching $ (n_{v} > n_{p}) $ 
vortex configurations having interstitially pinned vortices 
rather than multiple vortices per pinning site. 
For high matching fields, domains of different orientational order 
appear---including coexisting stripes and islands of approximately 
constant field (see Fig.~4).  These novel striped and 
island-type domains are separated by current-carrying winding paths 
made of $5$-- $7$--fold defect pairs.  These unexpected and complex
phases could be imaged by using, for instance, scanning Hall probes and 
Lorentz microscopy.

We have also shown in a quantitative manner that 
beyond the first matching field interstitially pinned vortices 
are much more mobile, leading to plastic flow of these weakly-pinned 
vortices around more strongly pinned ones trapped at defect sites. 
This result is consistent with recent experimental work with 
randomly placed columnar pinning \cite{Welp,Rosenbaum},
in which a sharp drop off in $M(H)$ was observed; this drop 
was interpreted as a consequence of the weak pinning felt by
interstitial vortices.
  
It is of interest to compare our results for commensurability  
at the first matching field to those predicted by the Bose glass 
theory \cite{Nelson}. 
Here, a Mott insulator-like state occurs in which 
the vortex lattice is locked into the pinning lattice for a  
certain field range, preventing further vortices from entering 
and creating a Meissner like rise in the magnetization. 
With a PAPS, we observe such a Mott insulator-like 
peak in the magnetization  near the first matching field; 
however, this peak is gradually lost as the pinning array 
is distorted [see Fig.~5(c)]. Experiments with random columnar 
pinning also do not observe  a peak in the magnetization at the first 
matching field \cite{Welp,Rosenbaum}. 
This suggests that because of the symmetry of the periodic pinning 
array all the  pinning sites are equally accessible so that the 
vortex lattice can easily lock into the pinning sites.  
Conversely, for a random distribution of defects, fluctuations in 
the locations of pinning sites can make some sites inaccessible 
due to screening of pins by repulsion from nearby occupied pinning
sites \cite{Olson}.  

For samples with a random distribution of pinning sites, fluctuations 
in the distribution can create regions of low pinning density which 
can give rise to easy-flow percolating paths for the vortices \cite{Olson} 
and thus reduce $J_{c}$. With a PAPS such fluctuations in pin density 
are minimized, thus maximizing $J_{c}$. 
Since $J_{c}$ is related to $M(H)$, our results show that the enhancement 
of the critical current for certain field regions is greater for periodic 
arrays of pinning sites than for random arrays.  

We thank J. Siegel for a critical reading of the manuscript.
SF was supported in part by the NSF under grant No.~DMR-92-22541. 
We acknowledge the use of the UM Center for Parallel Computing, 
partially funded by NSF grant No.~CDA-92-14296.

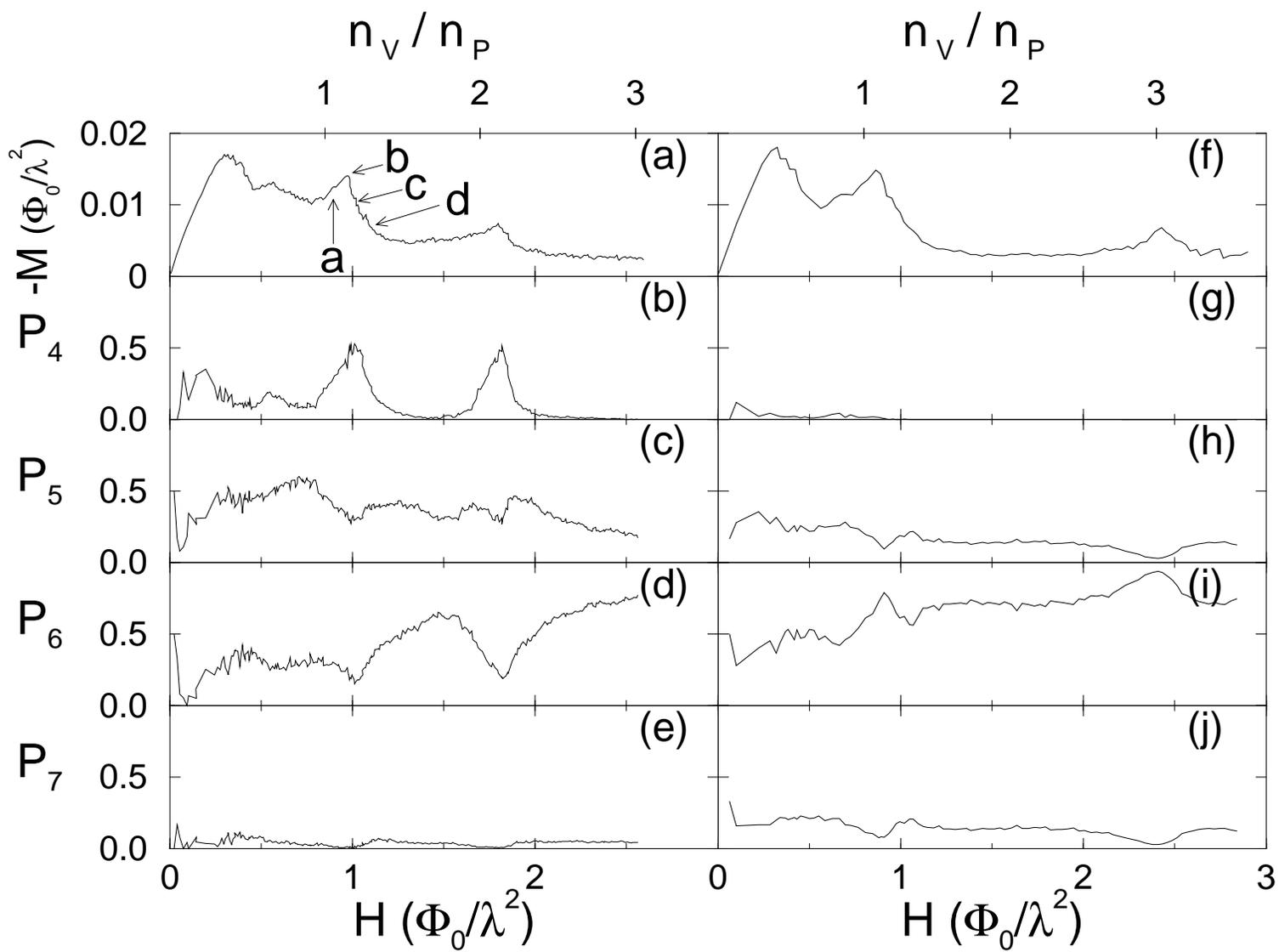
\begin{figure}
\caption[]{
Initial magnetization $M(H(t))$ for two samples with a pinned region 
of size $ 27\lambda \times 36 \lambda$ for a square (a) and triangular 
(f) periodic array of pinnings sites.  These are embedded in a 
$ 36\lambda\times36\lambda $ system in which the external field is brought 
from zero to a final value of $ H = 2.6\,\Phi_{0}/\lambda^{2}$,
corresponding to a total of $N_v=3400$ flux lines (2550 vortices in 
the pinned region) and $n_v \approx 2.6 /\lambda^2$.
All panels on the left (right) side refer to the square (triangular) 
periodic array of pinning sites. 
The pin densities are 
$ n_{p} = 0.86/\lambda^{2} $ ($N_p=834$)
for the square PAPS and  
$ n_{p} = 0.81/\lambda^{2} $ ($N_p=784$)
for the triangular PAPS.
$N_p$ is the total number of pinning sites.
For both of these samples $ f_{p} = 0.3f_{0} $, $\xi_{p} = 0.25\lambda $. 
For the square array, peaks in $M(H)$ near the 
$1/2$, $1$, and $2$ MFs can be observed in (a).
For the triangular array, peaks in $M(H)$ near the 
$1$ and $3$ MFs can be observed in (f).
The fraction of $n$-fold coordinated vortices $P_n(H)$ 
exhibit clear peaks in $P_4$, and dips in the other $P_n(H)$'s,
near the MFs for the square case.
The fraction of $n$-fold coordinated vortices $P_n(H)$ 
exhibit clear peaks in $P_6$, and dips in the other $P_n(H)$s,
near the MFs for the triangular case.
For increasing field in both PAPS, $P_6$ becomes close to one 
when $H > 2.5 \, \Phi_0/\lambda^2$; meanwhile all 
the other $P_n$'s decrease. 
}
\label{fig1}\end{figure}

\vskip +0.1in

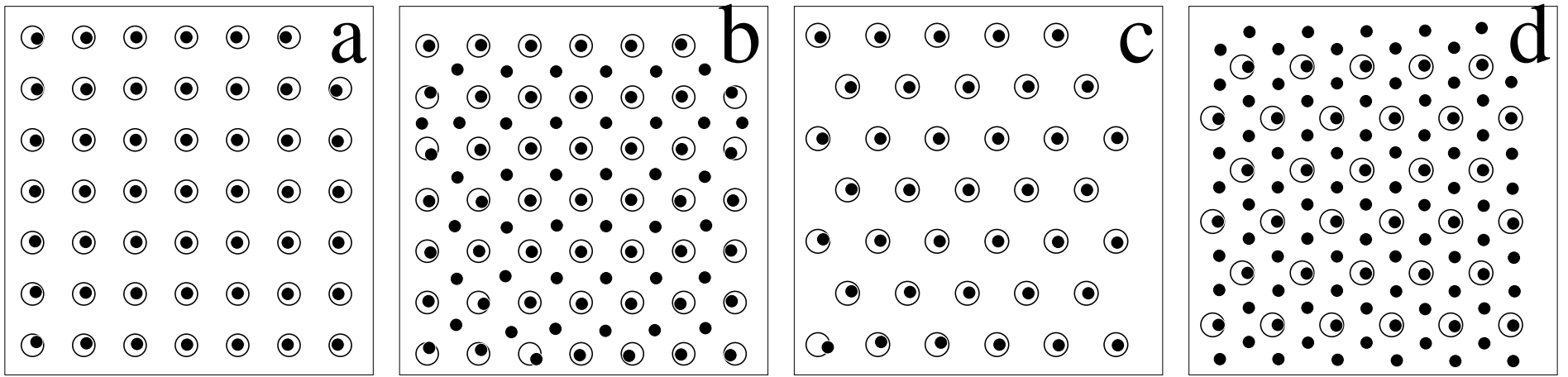
\begin{figure}
\caption[]{
Snapshots of a small part of the sample showing the actual 
location of vortices (solid dots) and pinning sites (open circles)
at the $M(H)$ peaks $n_v/n_p \approx 1, 2$, in (a,b), 
[$n_v/n_p \approx 1, 3$, in (c,d)] 
for the square [triangular] PAPS indicated in Fig.~1(a) [1(f)].
Note the interstitially pinned vortices in (b) and (d). 
}\label{fig2}\end{figure}  

\vskip +0.1in

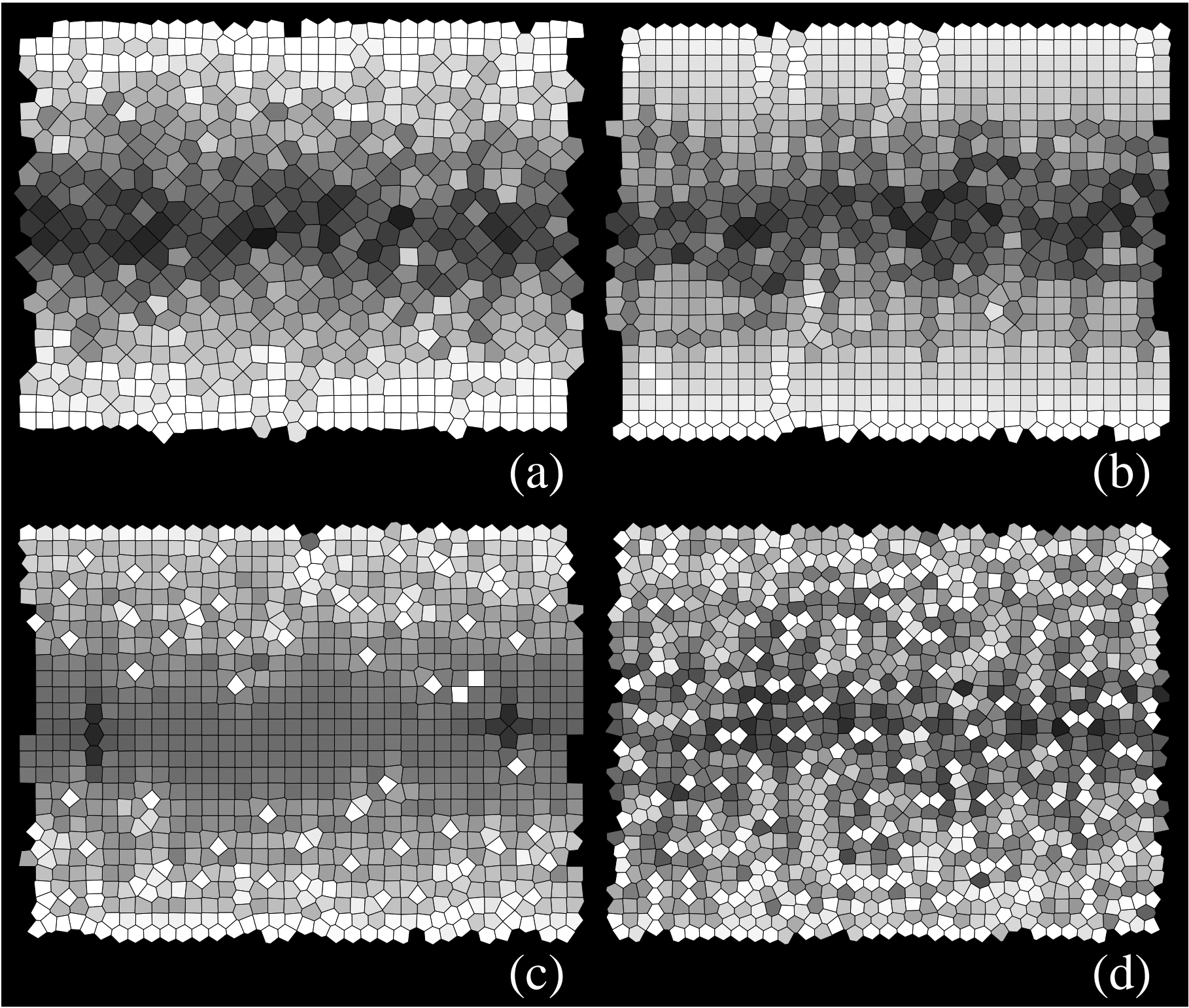
\begin{figure}
\caption[]{
(a--d) Voronoi (Wigner-Seitz) cell construction indicating the location 
of the vortices as they move through the first matching peak 
$n_{v}/n_{p} \approx 1$, for the square PAPS system described in Fig.~1(a). 
Larger (smaller) cells 
correspond to low (high) densities of flux lines and are 
indicated in dark (light) shading.
The sample has periodic boundary conditions on the left and right sides,
and the vortices penetrate at the top and bottom edges of the sample.
(a--d) correspond to locations on $M(H)$ indicated in Fig.~1(a) 
by the letters a--d.
Before the matching peak $n_{v}/n_{p} \approx 1$ [frame (a)],
and right at the matching peak [frame (b)], 
the PAPS and the VL are commensurate only near the edges 
of the sample, as indicated by the squares.  
(b) At the peak in $M(H)$, 
1D {\em string-like defects\/} which are perpendicular to the edges 
(or along $60^{\circ}$ angles for the triangular case) originate from 
the boundaries.  These gradually destroy the commensurability between 
the PAPS and the VL at the sample edges, shifting it towards the center 
of the sample [in (c)] when $H$ is increased.
In (c), just past the peak in $M(H)$, the smallest (white) 
diamond-shaped cells (which look like squares rotated $45^{\circ}$
from the horizontal) are mobile interstitial vortices whose plastic 
flow destroys the commensurability.  
These are also clearly present in (d), further down in $M(H)$, 
where commensurability effects dissappear.
This loss of commensurability can be quantified by observing in 
Fig.~1(c) the drastic drop in $P_4$ right after the commensurate peak.
}
\label{fig3}\end{figure}

\vskip +0.1in

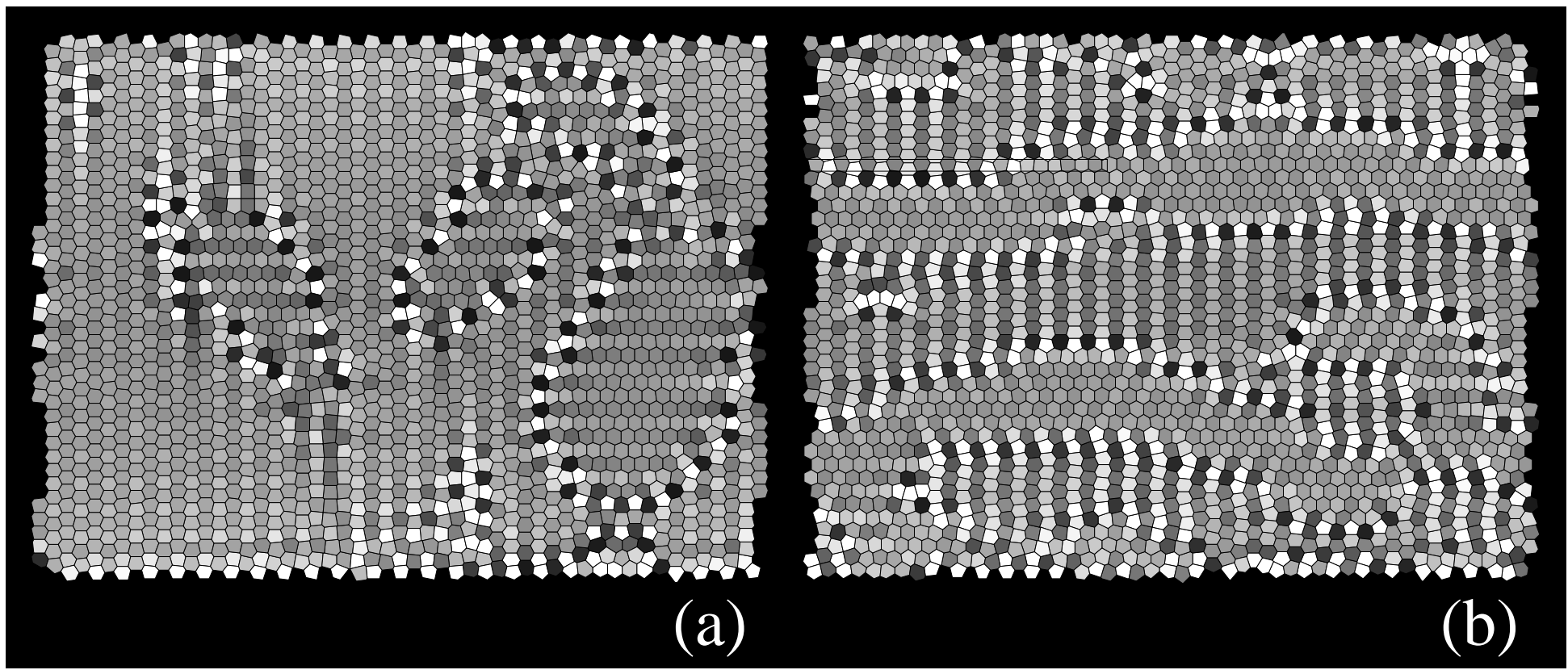
\begin{figure}
\protect\caption[]{
Vortex configurations for a square PAPS with a lower density of pins of
higher strength: $n_p = 0.49/\lambda^{2}$, $f_{p} = 1.4f_{0}$, 
$\xi_p = 0.25\lambda $, $N_p=475$, at (a) $ H=2.1\, \Phi_0/\lambda^{2} $ 
and (b) $ H=2.2\, \Phi_0 / \lambda^{2} $.
Vortex arrangements of different orientations form complex 
domains of various shapes including (a) islands and (b) stripes, 
both surrounded by {\it current-carrying winding strings\/} of 
$5$--$7$ disclination pairs.  These unexpected and striking 
flux domain patterns continuously evolve as a function of 
the applied field.
These results show that, in general, the current distribution is 
{\it not\/} terraced.  This is in contrast to predictions by 
static 1D models.  
We also note that the islands and stripes of (approximately) constant $B$, 
surrounded by curved domain-boundaries carrying large currents,
contain weak currents because in them 
\mbox{ $\nabla \times {\bf B} \approx 0$}.
}
\label{fig4}\end{figure}

\vskip -0.3in

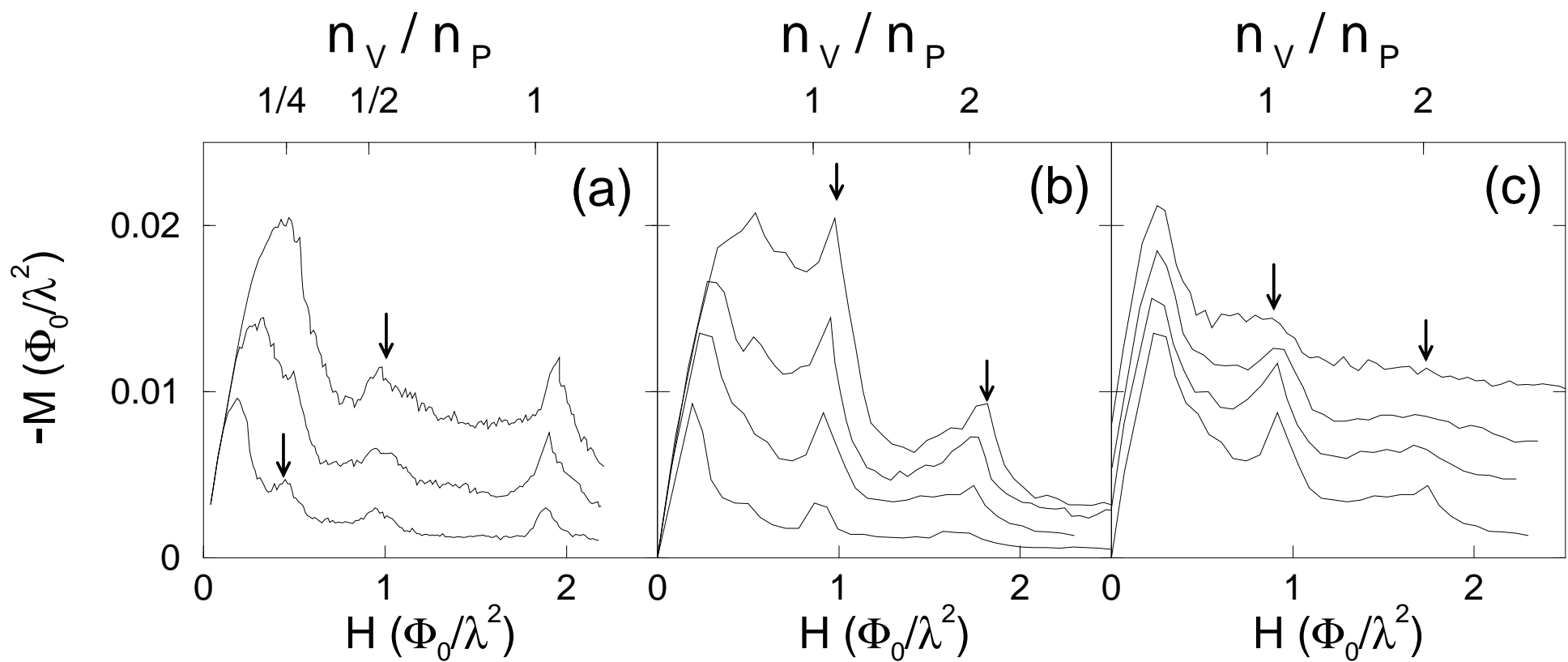
\begin{figure}
\protect\caption[]{
Magnetization versus field for samples with a square array of 
pinning sites.  
(a) $M(H)$ with submatching peaks at $1/4$ and $1/2$ for samples 
with weak pinning [$f_p / f_0 = 0.1$ (bottom curve), $0.2$, $0.3$ (top)], 
and $n_p$ about {\it twice\/} as large as in Fig.~1(a): 
$n_p=1.83/\lambda^2$, $N_p=1776$.
(b) shows $M(H)$ for increasing pinning strength and (c) shows 
$M(H)$ for increasing disorder in the location of the pins.
In (b) $n_p=0.86/\lambda^2$ and the pin strengths are varied: 
$ f_{p}/f_0 = 0.1 $ (bottom), $ 0.2 $, $ 0.3 $, and $ 0.4 $ (top).
In (c) $ f_p $ is fixed at $ 0.2 f_0 $, with disorder introduced 
gradually by randomly moving  
(with uniform probability) 
the pins from the initial square positions by distances up to
$\delta r$, for $\delta r = 0$ (bottom), $\lambda/8$, $\lambda/6$, 
and $\lambda/4$ (top).
All other parameters are the same as those used in Fig.~1(a).
For clarity, consecutive curves in (c) have been shifted vertically.
}
\label{fig5}\end{figure}


\vskip -0.3in

\begin{references} 

%\protect\vspace*{-0.3in}
%\vskip -0.2in

\bibitem{reviews}
For reviews, and  extensive lists of references, see:
M.G. Blamire, J.~Low Temp.~Phys.~{\bf 68}, 335 (1987);
A.N. Lykov, Adv.~Phys.~{\bf 42}, 263 (1993). 

\bibitem{L}
L.D.~Cooley {\it et al.}, Appl.~Phys.~Lett.~{\bf 64}, 1298 (1994).

\bibitem{Metulshko}
M.~Baert {\it et al.}, Europhys.~Lett.~{\bf 29} 157 (1995); 

\bibitem{Jonkeree}
M.~Baert {\it et al.}, Phys.~Rev.~Lett.~{\bf74}, 3269 (1995); 

\bibitem{Baert}
V.V.~Moshchalkov {\it et al.}, Jpn.~J.~App.~Phys.~{\bf 34}, 4559 (1995). 

\bibitem{Ovch}
A.~Bezryadin and B.~Pannetier, J.~Low Temp.~Phys.~{\bf 102}, 73 (1996);
A.~Bezryadin, Yu.~N.~Ovchinnikov, and B.~Pannetier, Phys.~Rev.~B {\bf 53},
 8553 (1996).
 
\bibitem{hunnekes}
C.~H\"unnekes     {\it et al.}, Phys.~Rev.~Lett.~{\bf 72}, 2271 (1994);
S.H.~Brongersma   {\it et al.}, {\it ibid.} {\bf 71}, 2319 (1993).
M.~Ziese {\it et al.}, Phys.~Rev.~B {\bf 53}, 8658 (1996).

\bibitem{JJ}
R.~Fehrenbacher, V.B.~Geshkenbein, and G.~Blatter
Phys.~Rev.~B {\bf 45}, 5450 (1992);
M.A.~Itzler and M.~Tinkham, {\it ibid.} % Phys.~Rev.~B 
{\bf 51}, 435  (1995). % , and references therein. 

\bibitem{networks}
See, e.g., 
M.A.~Iztler {\it et al.},
Phys.~Rev.~B %{\it ibid.} 
{\bf 42}, 8319 (1990);
Q.~Niu and F.~Nori, {\it ibid.} {\bf 39}, 2134 (1989). 

\bibitem{imaging}
L.N.~Vu, M.S.~Wistrom, and D.J.~Van Harlingen, 
Appl.~Phys.~Lett.~{\bf 63}, 1693 (1993); 
H.D.~Hallen {\it et al.}, Phys.~Rev.~Lett.~{\bf 71}, 3007 (1993);
K.~Runge and B.~Pannetier, Europhys.~Lett.~{\bf 24}, 737 (1993). 

\bibitem{oussena}
M.~Oussena {\it et al.}, Phys.~Rev.~Lett.~{\bf 72}, 3606 (1994). 

\bibitem{Grishin}
L.D.~Cooley and A.M.~Grishin, Phys.~Rev.~Lett.~{\bf 74}, 2788 (1995). 

\bibitem{Nelson}
D.R.~Nelson and V.M.~Vinokur, Phys.~Rev.~B {\bf 48}, 13060 (1993).

\bibitem{azbel} 
See, e.g., 
M.~Azbel, Zh.~Eksp.~Theor.~Fiz.~{\bf 46}, 929 (1964);
D.R.~Hofstatder, Phys.~Rev.~B {\bf 14}, 2239 (1976).

\bibitem{pinball}
D.~Weiss {\it et al.}, Phys.~Rev.~Lett.~{\bf 66}, 2790 (1991).

\bibitem{NL} E.A.~Jackson, {\it Nonlinear Dynamics} (Cambridge, 1991).

\bibitem{CDW} 
See, e.g., 
{\it Charge Density Waves in Solids}, L.P.~Gorkov and G.~Gr\"uner, eds.
(Elsevier, New York, 1989).

\bibitem{IK}
I.~Khalfin and B.~Shapiro, Physica {\bf 207C}, 359 (1993).

\bibitem{Leo}
L.~Radzihovsky, Phys.~Rev.~Lett {\bf 74}, 4919, 4923 (1995).

\bibitem{Majer}
M.~Konczykowski {\it et al.}, Physica {\bf C235-240}, 2965 (1994). 

\bibitem{Welp}
K.M.~Beauchamp {\it et al.}, Phys.~Rev.~Lett.~{\bf 75}, 3942 (1995)

\bibitem{Rosenbaum}
K.M.~Beauchamp {\it et al.}, Phys.~Rev.~B {\bf 52}, 13052 (1995).

\bibitem{Rosseel}
E.~Rosseel {\it et al.}, Phys.~Rev.~B {\bf 53}, R2983 (1996).

\bibitem{Science}
F.~Nori, Science {\bf 271}, 1373 (1996).

\bibitem{Tonomura} 
T.~Matsuda {\it et al.}, Science {\bf 271}, 1393 (1996).

\bibitem{Yaron}
U.~Yaron {\it et al.}, Nature {\bf 376}, 753 (1995), and work cited therein. 

\bibitem{Bean} C.P.~Bean, Rev.~Mod.~Phys.~{\bf 36}, 31 (1964). 

\bibitem{Kim}
Y.B.~Kim, C.F.~Hempstead, and A.R.~Strnad, Rev.~Mod.~Phys.~{\bf 36}, 43 (1964).

\bibitem{Richardson}
\mbox{R.A.~Richardson, O.~Pla, and F.~Nori, Phys.~Rev.~Lett.} {\bf 72}, 
1268 (1994). 

\bibitem{Reichhardt}
C.~Reichhardt, C.J.~Olson, J.~Groth, S.~Field, and F.~Nori, Phys.~Rev.~B, 
{\bf 52}, 10411 (1995). 

\bibitem{Olson}
C.~Reichhardt, C.J.~Olson, J.~Groth, S.~Field, and F.~Nori, Phys.~Rev.~B,
{\bf 53} R8898 (1996). 

\bibitem{simulations} See, e.g., 
O.~Pla and F.~Nori, Phys.~Rev.~Lett.~{\bf 67}, 919 (1991);
H.J.~Jensen, A.~Brass, A.-C.~Shi, and A.J.~Berlinsky,  
Phys.~Rev.~B {\bf 41}, 6394 (1990); and references therein.

\end{references}
\end{document}